\documentclass[a4paper]{article}
\usepackage{graphicx}
\usepackage{amsmath}

\begin{document}

\begin{center}
{\bf \Large
Order-disorder phase transition in a cliquey social network
}\\[5mm]

{\large
M. Wo{\l}oszyn$^{1,*}$,
D. Stauffer$^{1,2,\dag}$
and
K. Ku{\l}akowski$^{1,\ddag}$
}\\[3mm]

{\em
$^1$Faculty of Physics and Applied Computer Science,
AGH University of Science and Technology,
al. Mickiewicza 30, PL-30059 Krak\'ow, Euroland\\
$^2$Institute of Theoretical Physics,
Cologne University,
Z\"ulpicher Str. 77, D-50937 K\"oln, Euroland.
}

\bigskip
$^*${\tt woloszyn@agh.edu.pl},
$^\dag${\tt stauffer@thp.uni-koeln.de},
$^\ddag${\tt kulakowski@novell.ftj.agh.edu.pl}

\bigskip
\today
\end{center}

\begin{abstract}
We investigate the network model of community by Watts, Dodds and Newman (D. J. Watts et al.,
Science 296 (2002) 1302) as a hierarchy of groups, each of 5 individuals.
A homophily parameter
$\alpha$ controls the probability proportional to $\exp(-\alpha x)$ of selection of neighbours
against distance $x$.
The network nodes are endowed with spin-like variables $s_i = \pm 1$,
with Ising interaction $J>0$. The Glauber dynamics is used to investigate the
order-disorder transition. The transition temperature $T_c$ is close to 3.8 for
$\alpha < 0.0$ and it falls down to zero above this value. The result provides a mathematical
illustration of the social ability to a collective action {\it via} weak ties, as discussed
by Granovetter in 1973.

\end{abstract}

\noindent
{\em PACS numbers:} 89.65.s, \; 61.43.j

\noindent
{\em Keywords:}  sociophysics; hierarchy; phase transition

%% ############################################################################
\section{Introduction}
%% ############################################################################

To investigate the human society is more than necessary. However, the subject is probably
the most complex system we can imagine, whatever the definition of complexity could be. A
cooperation between the sociology and other sciences - including the statistical physics -
can be fruitful for our understanding of what is going around us. The science of social networks
seems to be a rewarding field for this activity \cite{nob,wat,sch,dor}. Although the physicists were
not inventors of basic ideas here, their empirical experience can be
useful at least for the mathematical modelling in social sciences. Moreover, it seems that purely
physical concepts like phase transitions can provide a parallel and complementary description
of phenomena observed by the sociologists. Such a description is also
a motivation for this research. Our aim is to investigate the social ability to organize,
as a function of the topology of a social ties network.

As it was stated by Granovetter \cite{gra} more than thirty years ago, the structure of social
ties can be a formal determinant in an explanation of the activity of a given community. Granovetter
wrote: "Imagine (...) a community completely partitioned into cliques, such that each person
is tied to every other in his clique and to none outside. Community organisation would be severely
inhibited." (\cite{gra}, p. 1373). As an example, the author provides "the Italian community
of Boston's West End (...) unable to even form an organisation to fight against the urban
revolution which ultimately destroyed it." Granovetter argued that new information is transported
mainly {\it via} distant connections (weak ties) between the cliques, and not within the cliques.

This compact description of a cliquey structure of a social network found recently a mathematical
realisation \cite{wdn}. There, the level of cliqueness was controlled along the following
receipt.
Initially, the community of $N$ individuals is virtually divided into $N/g$ small groups of $g$ individuals each.
These groups form the bottom level of a hierarchical structure defining the distances $x_{ij}$ between individuals $i,j=1,...,N$ as $x=1$ between the individuals in the same group, $x=2$ between the members of neighbouring groups, $x=4$ between the members of groups which form neighbouring groups
and so on.
A schematic view is shown in Fig. 1.
The virtual distances $x_{ij}$ are used to determine real links (the social ties) between the network nodes (individuals).
Namely, for each node $i$ its links to other nodes $j$ are drawn randomly, with the probability of a link between two nodes $i$ and $j$
depending on the distance $x_{ij}$ as $p_{ij} \propto \exp(-\alpha x_{ij})$.
The procedure is repeated until
a given number of neighbours $z=g-1$ on average is assured.
 In Ref. \cite{wdn}, the nodes were connected according to a set of a few of mutually intertwined hierarchies. Here we follow the original
 picture \cite{gra}, where only one hierarchy is present. This choice is determined by our aim to consider only ties due to one kind of activity, 
namely the political one. This activity leads to the social ability to organize, which is our interest in this work.

The topology of the network is
controlled by the parameter $\alpha$, called a homophily parameter in Ref. \cite{wdn}. For $\alpha =-\ln(2)$ every node
is selected with the same probability \cite{wdn}, then the system is just a random graph \cite{bollo}. (For a short introduction
to random graphs see for example \cite{err,new}). For $\alpha$
large and positive the links drawn reproduce the initial virtual separation of the community to small
groups. For $\alpha$ large and negative, far nodes are connected more likely. In Fig. 2 we present a graphical representation
of the connectivity matrix for a small system. Non-zero matrix elements are marked in black. As we see, for  $\alpha$
large and positive the only non-zero matrix elements are close to the diagonal and in fact, the matrix is decomposed into a set
of submatrices of the same size. This means that the system is split into a set of fully connected subgraphs, mutually separated.
For  $\alpha =-\ln(2)$ the connectivity matrix is random, with 4 units in each line on average. In this case, the graph is a random network
in the sense of Ref. \cite{bollo}. For $\alpha$ large and negative all links are placed between nodes which are distant in the sense
of remoteness in the hierarchy \cite{wdn}. The whole construction is exactly the same as in Ref. \cite{wdn} for the case of one hierarchy.
We note that the system is not similar to a linear chain. For $\alpha\approx$0.7, i.e. at the edge of splitting, the system can be approximated by 
two fairly dense clusters mutually connected by one or two links. These links are due to the largest possible distance, and therefore their
probability is the smallest.  On the other hand, we have checked that the number of isolated spins slightly decreases with $\alpha$.
The hierarchical structure of the network ensures the small world property, because the number of nodes is $N=5\times2^l$, where $l$ is 
the number of levels in the hierarchy, and the maximal distance is a linear function of $l$ \cite{wdn}.

\begin{figure}
\begin{center}
\includegraphics[scale=0.8]{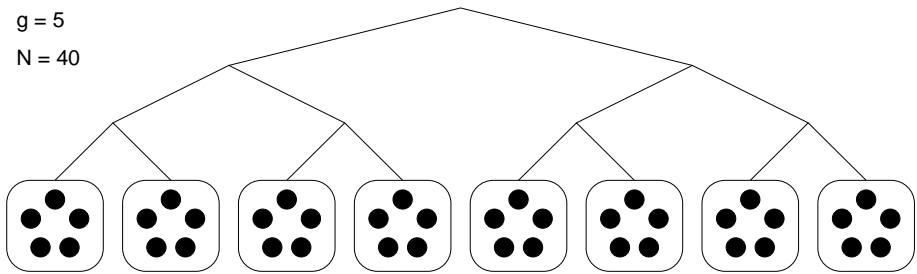}\\
\end{center}
\caption{A schematic view of the system for $g$=5.}
\end{figure}

\begin{figure}
\begin{center}
\includegraphics[scale=0.7]{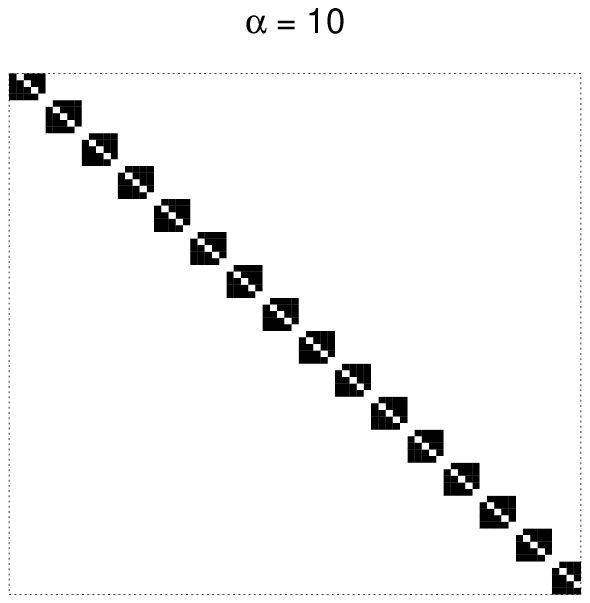}\\
\includegraphics[scale=0.7]{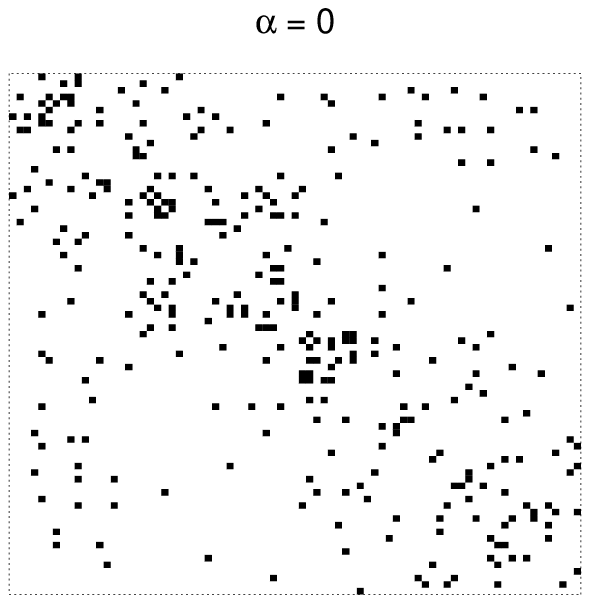}\\
\includegraphics[scale=0.7]{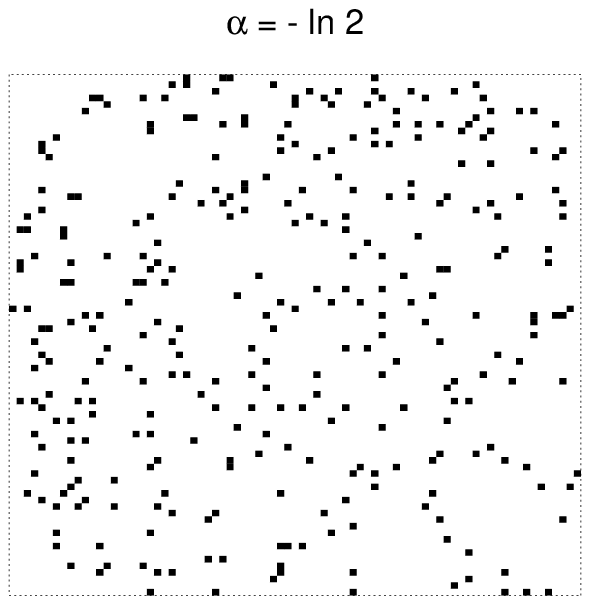}\\
\includegraphics[scale=0.7]{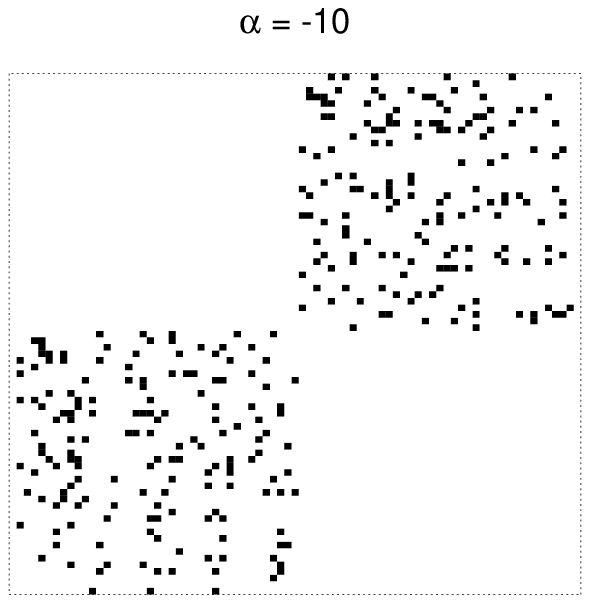}
\end{center}
\caption{Non-zero elements of the connectivity matrix for $g$=5, $N$=80 $\alpha = 10.0, \; 0.0, \; -\ln(2), \; -10.0$, from top to bottom.}
\end{figure}

Here we add one more ingredient to the model. A spin $s_i$ is assigned to each node, and an interaction
energy $J$ to each link. The binary approximation of opinions has got some support in the threshold model of social behaviour \cite{thres}.
The energy $J$ is the same for each link, and it favors the same sign of spins of
neighbouring nodes. In this way the social system is translated into a magnet with the topology close to
the one suggested by Granovetter. As with a magnet, we can ask if any kind of a phase transition is possible \cite{ale,her,tor}
where the spins order below some level of thermal noise to have mostly the same orientation. This
phase transition, if it is present in the magnetic system, serves here as a parallel to measure the
ability of the social system to a collective action. Oppositely, a lack of the transition can be interpreted
as an indication that the network cannot behave in a coordinated way. Using this model, we do not state that
the magnetic interaction is in any sense similar to the interpersonal interaction. We only assume that an
influence of the topology of the social ties on the social collectivity can be reproduced to some extent
by the influence of the network topology on a collective state, with the latter measured by a scalar spin variable.

We should add that the connection of magnetic phase transitions to collective social phenomena is by no means new. Several authors evaluated
an influence of topology of small-world and scale-free networks on the existence and universality class of the ferro-paramagnetic phase transition 
of Ising spins, placed at the network nodes \cite{dgm,klem,sanch,grab,lop}. The authors of these papers (and presumably several others) referred
to the magnetic phase transition as to an equivalent to collective social phenomena: the cultural globalisation \cite{klem}, the formation of cultural domains \cite{sanch}, modelling of social opinion including mass-media \cite{grab} and so on. Other references to the Ising spins on disordered networks can be found e.g. in \cite{new}. The list of authors who considered models of the social network is much longer;
in fact, the subject is well established in sociology \cite{sn}. Also, the binary state of an Ising spin make it a convenient starting point to numerous
applications in the game theory, where the concepts of energy and temperature have only indirect equivalents \cite{game}.

\section{Calculations and results}

The calculations are performed for networks of $N$ nodes, from $N=640$ to 10240. The time of calculation was $10^5$ steps, all spins updated at one step.
The obtained results were averaged over last $5\times 10^4$ steps.

As shown in Fig.2, the structure of the obtained network depends on the parameter $\alpha$. To capture the details of this dependence,
we calculated numerically the size $n_{\mathrm{max}}$ of the largest component of the network and the clustering coefficient $C$ against $\alpha$.
The results are shown in Figs. 3, 4. For our purposes, the properties of the network are of interest mostly at the edge of its breaking
into pieces, where the ties connecting local groups are indeed weak. We read from Fig. 3 that this is the region where $0<\alpha<0.7$.
The clustering coefficient $C_i$ of node $i$ is usually defined as the number of existing links between nodes connected to $i$
divided by the maximal possible number of these links \cite{new}, whereas $C$ is an average of $C_i$ over all nodes. (Some subtleties of
this definition are discussed in Ref. \cite{corrcoe}.) As shown in Fig. 4, the clustering coefficient $C$ increases in the above range of
$\alpha$ from almost zero to about 0.2.

\begin{figure}
\begin{center}
\includegraphics[scale=0.8]{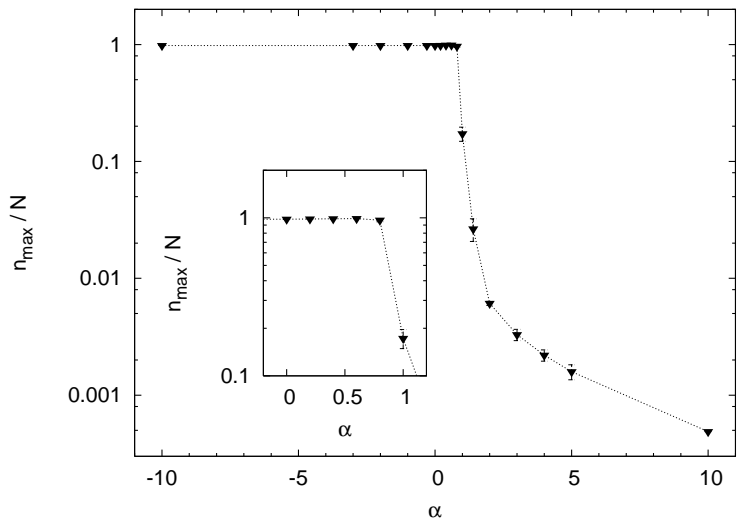}\\
\end{center}
\caption{Amount of nodes $n_{\mathrm{max}}/N$ in the largest connected part of network against the coefficient $\alpha$. The result is an 
average over 4 samples of 10240 nodes. The error bars are marked. The curve slightly increases 
near $\alpha \approx$0.6, and sharply decreases above $\alpha$=0.7, splitting into two clusters of approximately equal sizes.}
\end{figure}

\begin{figure}
\begin{center}
\includegraphics[scale=0.8]{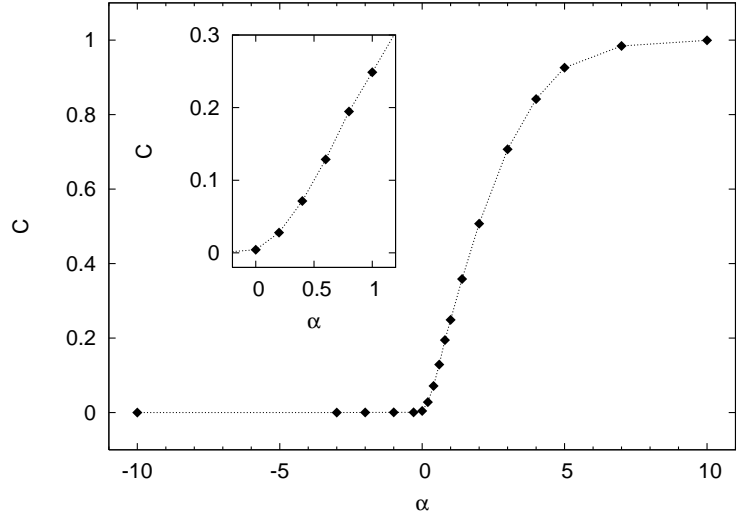}\\
\end{center}
\caption{The clustering coefficient $C$ against the coefficient $\alpha$. The result is an 
average over 4 samples of 10240 nodes. The error bars are of size of the symbols used.}
\end{figure}

In Fig. 5 we show the thermal dependence of the magnetisation for some values of $\alpha$. These and other results allow to calculate
the critical temperature $T_c$ versus $\alpha$ - these results are shown in Fig. 6. The result for $\alpha=-\ln(2)$ is approximately
confirmed for $N=10^6$ nodes \cite{aman}. Moreover, in this case the structure is equivalent to the Erd{\H o}s--R\'enyi random network 
\cite{bol}. In this case, the exact analytical solution \cite{dgm} gives $T_c=3.91$. The dependence $T_c(\alpha)$ was calculated
also for the network sizes $N$=2560 and 5120. The results indicate that the obtained dependence falls down more sharply as $N$ increases.
 For $\alpha > 1.2$ the time of calculations
necessary to get thermal equilibrium becomes very large and we have no reliable results. However, it is clear that the critical temperature
of small groups of $g=5$ nodes is zero. This supposition is supported by the numerical results of the time of relaxation to equilibrium
for these clusters, shown in Fig. 7. The data obtained with this method are marked with crosses in Fig. 6.

\begin{figure}
\begin{center}
\includegraphics[scale=0.8]{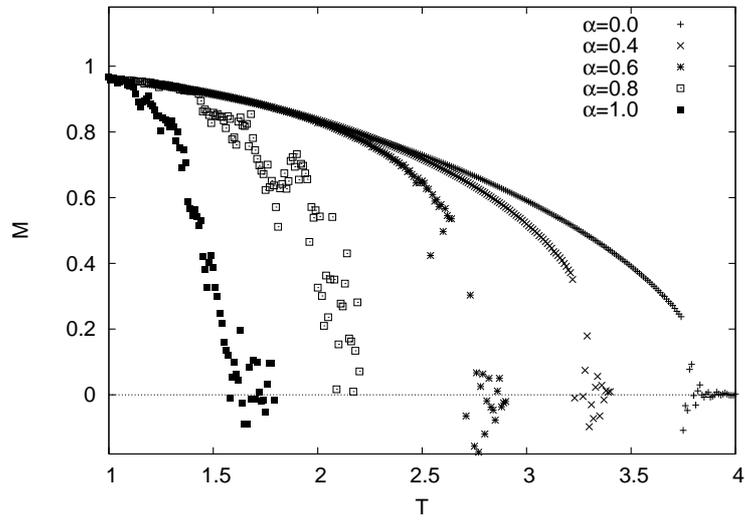}\\
\end{center}
\caption{Magnetisation $M$ against temperature $T$ for some selected values of $\alpha$.}
\end{figure}

\begin{figure}
\begin{center}
\includegraphics[scale=0.8]{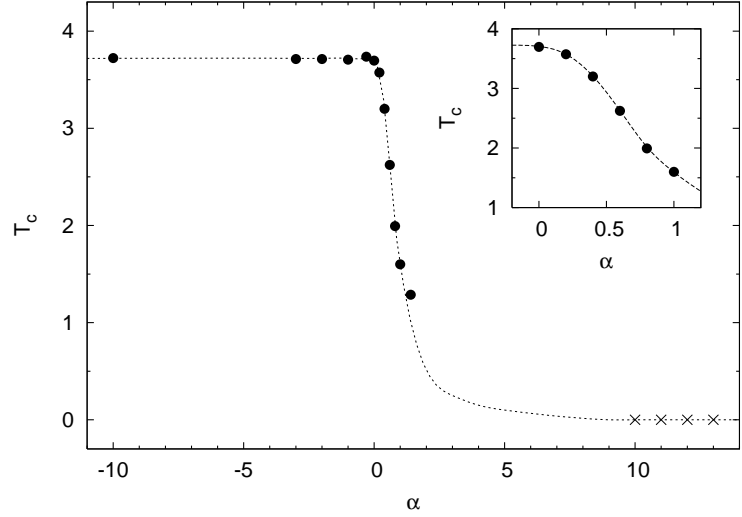}\\
\end{center}
\caption{The critical temperature $T_c$ against the coefficient $\alpha$. The result is an 
average over 4 samples of 10240 nodes. The error bars are of size of the symbols used. The dotted line is a guide to the eye.}
\end{figure}

\begin{figure}
\begin{center}
\includegraphics[scale=0.8]{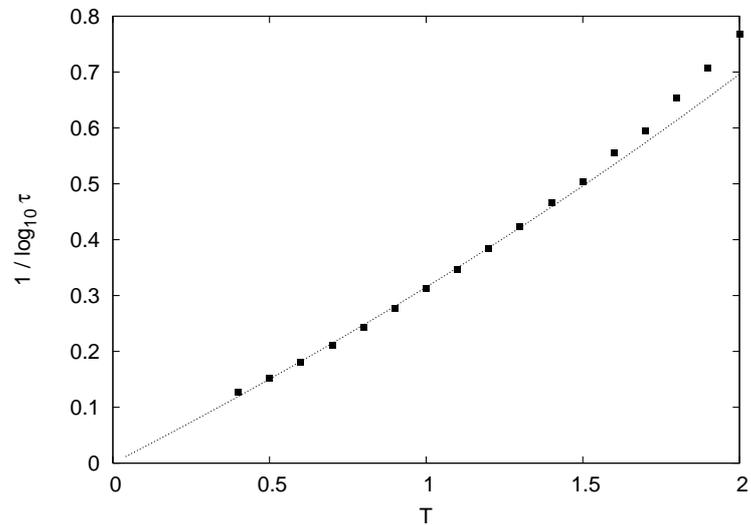}\\
\end{center}
\caption{The relaxation time $\tau$ for $\alpha$=10.0, for various system sizes $N$. The line is $\tau = 0.5 \exp(8/T)$. The data
were obtained for $N$ up to $25 \times 10^4$ nodes.}
\end{figure}

\section{Discussion}

Our numerical results clearly indicate, that the critical temperature of the system sharply decreases with the homophily
parameter $\alpha$. This result is easy to interpret when we deal with the case of $\alpha>0.7$, when the network is divided into
parts. However, the critical temperature starts to decrease already above $\alpha=0$, where the network is still connected. This means
that the links between separated groups become too scarce to maintain the magnetic ordering. One can imagine a tree of small clusters
almost fully connected inside, but with only one or two links to other clusters. The ability of such a system to show the long-range
magnetic ordering is not much larger than in the case of one-dimensional chain of spins, where $T_c=0$. In this sense the weak ties
between the clusters become too weak when $\alpha>0$. We note that in our system, all links are of the same strength. However, 'weak ties'
in the sense of Granovetter \cite{gra} mean not their strength, measured by e.g. the intensity of contact, but rather their unique function
to provide contact between far groups, otherwise separated. To refer again to the original text, "removal of the average weak tie would 
do more "damage" to transmission probabilities than would that of the average strong one." \cite{gra}.

Assuming that the interpretation of the phase transition is sociologically meaningful, we can state that
our numerical result agrees with the qualitative prediction of Granovetter, made in 1973. As long as the
connections between the small groups are too sparse, the system as a whole does not show any collective
behaviour. We note that the number of links in the network does not vary with the homophily parameter $\alpha$. It
is only their distribution which changes the system behaviour. Obviously, we have no arguments to defend
particular elements of the model, like the number of states
of one node (which is two), or the homogeneous character of the node-node interaction (the same for each tie),
or the tie symmetry (the same in both directions) etc. All these model ingredients should be treated as particular
and they can vary from one approach to another. On the contrary, as we deduce from the universality hypothesis,
the phase transition itself does depend on the number of components of the order parameter \cite{sta}. The assumption
on the Ising model is nontrivial, but remains arbitrary. The argument is that the model is the simplest known.
It would be of interest to check our results for more sophisticated descriptions of the social interactions,
as the models of Sznajd~\cite{szn}, Deffuant~\cite{def} or Krause-Hegselmann~\cite{krh}.

Still there is a gap between the sociological description of the structure of the social network and the global characteristics
used in the sociophysics. The former is much more detailed and oriented to look for differences between particular cases,
whilst the second is theoretical and tends to capture general features, neglecting details. An example of the sociological approach
is Ref. \cite{louch}; there, homophily is found to enhance the so-called transitivity in personal networks of strong ties. As the transitivity
can be measured by the clusterisation coefficient, Ref. \cite{louch} can be seen as a complementary to the paper of Granovetter \cite{gra},
which was the starting point of our study.

Concluding, it is not the critical value of the homophily parameter $\alpha$ which is relevant for the
sociological interpretation, because this critical value depends on all the above mentioned details. What
is - or can be - of importance is that this critical value exists. The task, how to model a collective state
in a social system, remains open. We can imagine that exceeding the critical value of some payout,
common for a given community, could trigger off a collective action, enhanced then by a mutual interaction.
Attempts of this kind of description, with the application of the mean field theory, are classical in sociology \cite{thres}
as well as in sociophysics \cite{gal}. The result of the present work assures that the effectiveness of such a social interaction depends on
the topology of the social network. The same approach can be applied - and was applied - to other models of the social structure
(\cite{was,gir,boymi}).

\bigskip

%% ============================================================================
\noindent
{\bf Acknowledgements.} The work was partially supported by COST Action P-10 "Physics of Risk".
%% ============================================================================

%% ############################################################################

\end{document}